\documentclass[aps,prapplied,12pt,superscriptaddress,notitlepage]{revtex4-2} 
\usepackage[utf8]{inputenc}
\usepackage{amsmath, amssymb, bm}      % Math packages
\usepackage{graphicx}                   % Graphics
\usepackage{wrapfig}                    % Wrapping text around figures
\usepackage{array}                       % Array and tabular enhancements
\usepackage{booktabs}                   % Better tables
\usepackage{setspace}                   % Line spacing
\usepackage{placeins}                   % Float control
\usepackage{chngcntr}                   %change fig cntr for Supp.
\usepackage{braket}
\usepackage{hyperref}
\UseRawInputEncoding
\usepackage{xr}                         % External references

\usepackage{tikz}                       % TikZ package
\usetikzlibrary{trees}                  % TikZ libraries
\begin{document}
\title{Nonreciprocal Thermophotonic Cooling}

\author{Daniel Cui}\affiliation{Department of Materials Science and Engineering, University of California, Los Angeles, Los Angeles, CA 90095, USA}
\author{Parthiban Santhanam}\affiliation{Department of Materials Science and Engineering, University of California, Los Angeles, Los Angeles, CA 90095, USA}
\author{Aaswath P. Raman}
\email{aaswath@ucla.edu}

\affiliation{Department of Materials Science and Engineering, University of California, Los Angeles, Los Angeles, CA 90095, USA}
\affiliation{California NanoSystems Institute, University of California, Los Angeles, Los Angeles, CA 90095, USA}

\begin{abstract}

Solid-state cooling via electroluminescent emission from light-emitting diodes is a promising alternative to thermoelectric and vapor-compression refrigeration, but practical performance remains limited by nonradiative losses and unfavorable tradeoffs between efficiency and cooling power. Thermophotonic (TPX) architectures partially address this by recycling PV-generated power back to the LED, improving the coefficient of performance (COP) but introducing a parasitic backward photon flux from the PV that reduces the cooling power density. Here we show that this tradeoff can be circumvented by inserting a nonreciprocal semi-transparent intermediate layer that violates Kirchhoff's law of thermal radiation. The layer permits unity transmission from the LED to the PV while fully absorbing the backward PV flux, functioning as a radiative heat shield that re-emits toward the LED at a lower intermediate temperature. In the idealized limit for $\Delta T = 50$~K between the hot and cold side, the nonreciprocal filter improves the cooling power density by nearly an order of magnitude over the unfiltered TPX case while preserving the COP benefit, while a reciprocal filter provides no improvement. Incorporating Shockley-Read-Hall and Auger recombination into GaAs and InP-based LED device models, we find enhancements of approximately 50\% in both cooling power density and COP persisting across temperature differences from $\Delta T = 50$~K to 100~K. These results highlight the potential importance of electromagnetic nonreciprocity in improving the real-world performance of thermophotonic cooling devices.

\end{abstract}
\maketitle 

\section{Introduction}

Light-emitting diodes can be used for solid-state refrigeration when biased below their bandgap energy \cite{tauc_share_1957,santhanam_thermoelectrically_2012,radevici_thermophotonic_2019}. As an LED emits photons with energy greater than its electrical bias energy, the difference is supplied by drawing heat from the semiconductor lattice through phonons \cite{pipe_k_bias_dep_Peltier2002,sadi_thermophotonic_2020}, producing a cooling effect known as electroluminescent cooling (ELC). This mechanism is attractive as a potential solid-state alternative to thermoelectric coolers, offering the possibility of high efficiency cooling without moving parts or working fluids. In a single-junction ELC device, the emitted photons are typically absorbed by a passive blackbody absorber, and the net cooling power is determined by the balance between the heat extracted per emitted photon and the losses due to nonradiative recombination. In practice, the internal quantum efficiency is degraded by Shockley-Read-Hall (SRH) and Auger recombination processes, both of which become increasingly dominant at the sub-bandgap bias voltages where cooling occurs \cite{santhanam_room_2013}. As a result, experimental demonstrations of net ELC have remained limited, with only modest cooling powers achieved at room temperature to date.

Thermophotonic (TPX) architectures offer a route to improved system-level performance by replacing the passive blackbody absorber with a photovoltaic (PV) cell, which converts a portion of the emitted luminescence back into electrical power that is recycled to the LED \cite{radevici_thermophotonic_2019,chen_high-performance_2017, xiao_electroluminescent_2018, sadi_electroluminescent_2019}. This power recovery reduces the net electrical input required to sustain a given photon flux, thereby enhancing the overall cooling efficiency as measured by the coefficient of performance (COP). The TPX concept has attracted growing interest for applications in near-field and far-field solid-state cooling \cite{habibi_zero_2026,ng_farfield_2024,chatelet_performances_2025}, and recent modeling efforts have explored the roles of photon recycling \cite{Sutter_photrecyc_2017}, sub-bandgap photon management \cite{chen_suppressing_2015}, and thermal resistance in determining overall device performance \cite{zhao_self_2019}. However, the introduction of a PV cell also creates a new loss channel: thermal radiation from the PV cell back toward the cold side partially offsets the cooling effect of the LED, and this parasitic heat flux grows with the PV cell's operating temperature. As a consequence, improving the COP in a TPX system typically comes at the expense of reduced cooling power density relative to conventional ELC with a blackbody absorber, presenting a fundamental tradeoff that has motivated interest in new physical mechanisms to improve both metrics simultaneously.

Recent experimental and theoretical work has shown that Kirchhoff's law of thermal radiation which dictates  the equality of spectral emissivity and absorptivity can be broken under certain conditions. Experiments in highly doped semiconductors subject to strong magnetic fields have demonstrated broadband nonreciprocal thermal emission \cite{shayegan_broadband_2024,shayegan_nonreciprocal_2022,zhu_strong_nr_2025}, while theoretical studies have explored leveraging the intrinsic magnetic behavior of Weyl semimetals to achieve similar violations without external fields \cite{zhao_axion-field-enabled_2020,tsurimaki_large_2020,butler_wsm_2023}. These findings have spurred proposals for device applications that exploit thermal nonreciprocity. In particular, violating Kirchhoff's law at the interfaces of semi-transparent media has been shown to enable unidirectional radiative transport analogous to an optical isolator \cite{park_violating_2021,asadchy_WSM_opticalisolator_2020}, a concept that has been shown theoretically to enhance thermophotovoltaic (TPV) efficiency \cite{park_nonreciprocal_2022,zhao_high-performance_2017,Ghalekohneh_Nonrecip_Solar_TPV2022} and to approach the Landsberg limit in multijunction solar cells \cite{park_reaching_2022,park_does_2022}. However, employing nonreciprocity for thermophotonic and electroluminescent cooling has not been explored thus far.

In this Article, we show that violating Kirchhoff's law of thermal radiation in a semi-transparent intermediate layer can simultaneously enhance both the COP and cooling power density in a thermophotonic cooling system, circumventing the conventional tradeoff between these two figures of merit. The intermediate nonreciprocal layer functions as a radiative heat shield, selectively blocking a portion of the parasitic heat flux from the PV cell back to the cold side. We first demonstrate this enhanced performance in an idealized system free of material nonidealities, and then show that improvements of approximately 50\% in both cooling power density and COP persist under realistic material parameters for GaAs and InP-based device models. Our results provide both a theoretical framework and promising results for the role of nonreciprocal thermal emission in next-generation electroluminescent cooling-based devices.

\section{Nonreciprocity for Electroluminescent Cooling in the Ideal Limit}

\subsection{Electroluminescent Cooling Framework}

\begin{figure}[h!]
    \centering
    \includegraphics[width = 5.5in]{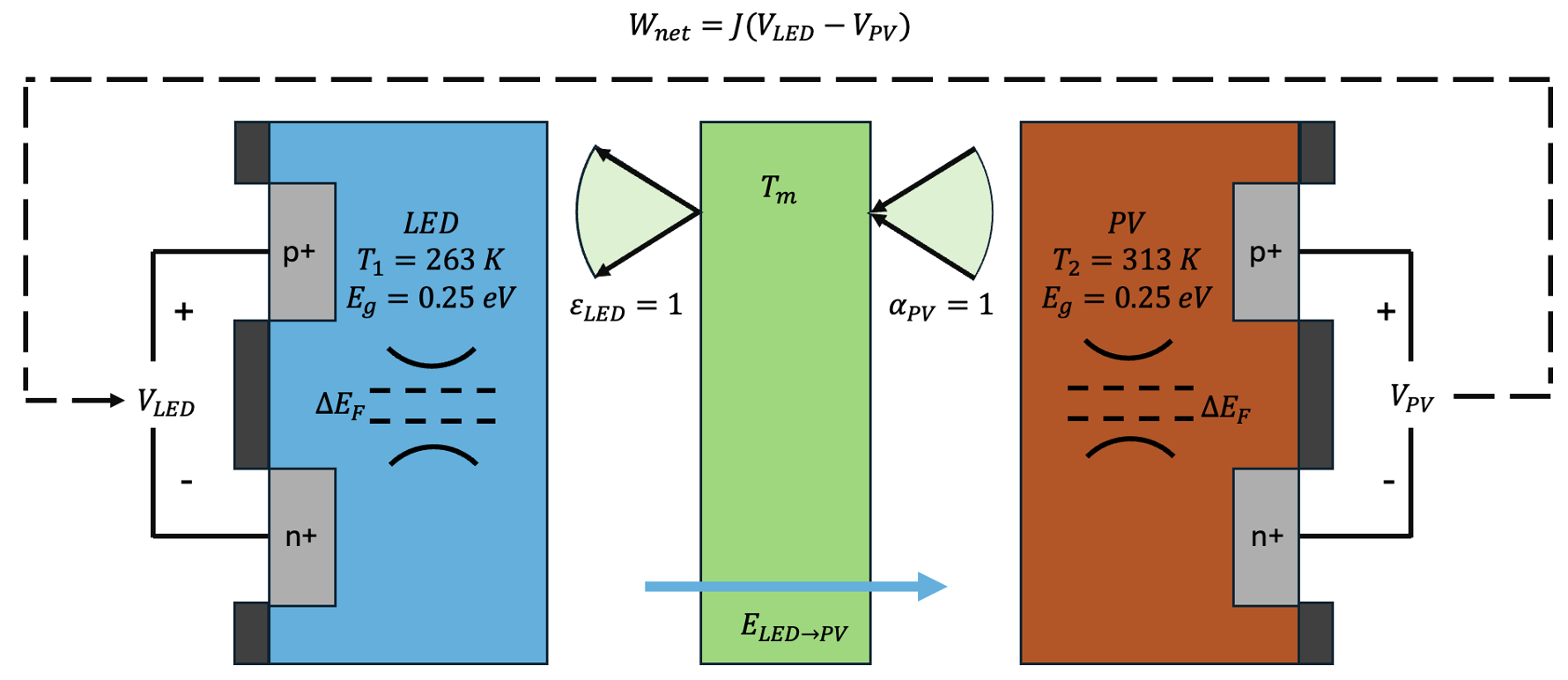}
    \caption{Schematic of thermophotonic cooling where the hot side is a biased PV cell. The LED, while in forward bias, emits a photon flux while the absorber emits a backward photon flux from its own thermal radiation. A portion of the work done by the PV, generated by the absorbed photon flux from the LED, is fed back into the LED to improve the COP. A nonreciprocal intermediate filter is situated in between that allows unity transmission of photons from the LED to the PV. In the opposite direction, transmission from the PV is blocked, but the filter still absorbs radiation from the PV and emits radiation toward the LED. It is important to note that Kirchhoff's law of thermal radiation is maximally violated for a narrow angular range to yield unity emissivity on the side facing the LED and unity absorptivity on the side facing the PV.}
    \label{fig:schematics}
\end{figure}

We consider a semi-infinite LED held at a cold temperature $T_1$ that emits above-bandgap photons to a semi-infinite hot absorber held at $T_2$ by a hot bath (Fig.~\ref{fig:schematics}). The spectral photon flux is given by
\begin{align}
    \phi(E,\mu,T) = \frac{2\pi}{h^3 c^2} \frac{E^2}{e^{(E-\mu)/k_B T}-1}
\end{align}
where $\mu = qV$ is the photon chemical potential set by the applied bias \cite{p_wurfel_chemical_1982,zhao_photon_chempotential2020,ross_therm_photochemical_1967}. The LED emits with chemical potential $\mu_1 = qV_1$, while the absorber side has chemical potential $\mu_2$ (with $\mu_2 = 0$ for a passive blackbody absorber and $\mu_2 = qV_2$ for a biased PV cell). We assume that the semiconductor materials possess sharp band edges and emit and absorb only above the bandgap energy $E_g$, that the transmission coefficient is unity, and that the radiation exchange occurs in the far field where near-field enhancement effects can be neglected \cite{chen_nearfield_enhance_2015,fiorino_nanogap_2018,zhu_near-field_2019}.
When operating in the electroluminescent cooling regime, radiative recombination of electrons and holes in the PN junction extracts heat from phonons in the semiconductor lattice and transfers it into the luminescent photon flux, carrying away an entropy flux from the LED. The current density, input power density, net luminescent heat flux from the cold to hot side, and cooling power density are respectively
\begin{align}
    J &= \frac{2\pi q}{h^3 c^2} \int^{\infty}_{E_g} \left[\frac{E^2}{e^{(E-qV_1)/kT_1}-1} - \frac{E^2}{e^{(E-\mu_2)/kT_2}-1}\right] dE \\[6pt]
    W_{\mathrm{LED}} &= J V_1 \\[6pt]
    Q_H &= \frac{2\pi}{h^3c^2} \int^{\infty}_{E_g} E \left[\frac{E^2}{e^{(E-qV_1)/kT_1}-1} - \frac{E^2}{e^{(E-\mu_2)/kT_2}-1}\right] dE \\[6pt]
    Q_L &= Q_H - W_{\mathrm{LED}} = \frac{2\pi}{h^3 c^2} \int^{\infty}_{E_g} (E-qV_1)\left[\frac{E^2}{e^{(E-qV_1)/kT_1}-1} - \frac{E^2}{e^{(E-\mu_2)/kT_2}-1}\right] dE
\end{align}

The ultimate efficiency of this setup is obtained when the net entropy generation vanishes, i.e., $Q_L/T_1 = Q_H/T_2$. Solving for the coefficient of performance $\mathrm{COP} = Q_L / W$ yields
\begin{align}
    \mathrm{COP}_{\mathrm{Carnot}} = \frac{T_1}{T_2 - T_1}
\end{align}
In the narrow-bandwidth limit around $E_g$ (where the photon flux is concentrated), the entropy balance condition $Q_L = (T_1/T_2)\, Q_H$ can be used to show that the threshold voltage for cooling onset is
\begin{align}
    V_t = \frac{E_g}{q}\left(1 - \frac{T_1}{T_2}\right)
\end{align}
At this voltage, the Bose--Einstein occupation factors on the LED and absorber sides become equal at $E = E_g$, which forces $J = 0$ and consequently $Q_L = 0$. This highlights the well-known fundamental tradeoff between COP and cooling power density: operating at Carnot COP at the threshold voltage drives the cooling power to zero \cite{Benenti_ThermBounds_TRS_2011,Shiraishi_tradeoff_powereff_2016,Pietzonka_steadystate_tradeoff_2018}.

At bias voltages above $V_t$, the cooling power density can be improved by introducing a band-pass filter that transmits only photons within an energy window $[E_g, E_c]$, blocking higher-energy photons that contribute to net heating. Such spectral filtering has been proposed to enhance the efficiencies of multijunction solar cells and ELC \cite{brown_limiting_2002,park_multijunction_2024,marti_limiting_1996}. The cutoff energy $E_c$ is determined by the condition that the photon fluxes from the hot and cold sides are balanced, i.e., $\phi(E_c, qV_1, T_1) = \phi(E_c, \mu_2, T_2)$, which in the Boltzmann approximation yields
\begin{align}
    \frac{E_c - qV_1}{kT_1} &= \frac{E_c - \mu_2}{kT_2} \nonumber\\
    E_c &= \frac{qV_1 - \mu_2\, T_1/T_2}{1 - T_1/T_2}
    \label{eq:cutoff_general}
\end{align}
For a passive blackbody absorber ($\mu_2 = 0$), this reduces to $E_c = qV_1/(1-T_1/T_2)$, while for a biased PV ($\mu_2 = qV_2$) the full expression in Eq.~\eqref{eq:cutoff_general} applies.

\subsection{Reciprocal Intermediate Layer}

We first examine the optical properties of a reciprocal intermediate layer. Outside the pass band $[E_g, E_c]$, the filter is set to be perfectly reflecting ($R_{\mathrm{LED}} = R_{\mathrm{PV}} = 1$, $T_{\mathrm{LED}} = T_{\mathrm{PV}} = 0$), which by energy conservation forces $\alpha = \epsilon = 0$ on both sides. Within $[E_g, E_c]$, we assume unity transmission from the LED side ($t_{\mathrm{LED}} = 1$). Energy conservation then requires $\alpha_{\mathrm{LED}} + R_{\mathrm{LED}} + t_{\mathrm{LED}} = 1$, giving $\alpha_{\mathrm{LED}} = R_{\mathrm{LED}} = 0$. By Kirchhoff's law, $\epsilon_{\mathrm{LED}} = \alpha_{\mathrm{LED}} = 0$, and by reciprocity $t_{\mathrm{PV}} = t_{\mathrm{LED}} = 1$, so the same holds on the PV side. Consequently, within $[E_g, E_c]$ the reciprocal filter exchanges no energy with either the LED or the absorber, and its temperature can be ignored. The performance metrics $Q_L$, $Q_H$, and $W$ take the same form as Eqs.~(2)--(5) with the integration limits replaced by $[E_g, E_c]$:
\begin{align}
    Q_L &= \frac{2\pi}{h^3 c^2} \int^{E_c}_{E_g} (E-qV_1)\left[\frac{E^2}{e^{(E-qV_1)/kT_1}-1} - \frac{E^2}{e^{(E-\mu_2)/kT_2}-1}\right] dE \nonumber\\
    Q_H &= \frac{2\pi}{h^3c^2} \int^{E_c}_{E_g} E \left[\frac{E^2}{e^{(E-qV_1)/kT_1}-1} - \frac{E^2}{e^{(E-\mu_2)/kT_2}-1}\right] dE \nonumber\\
    W_{\mathrm{in}} &= \frac{2\pi q V_1}{h^3 c^2} \int^{E_c}_{E_g} \left[\frac{E^2}{e^{(E-qV_1)/kT_1}-1} - \frac{E^2}{e^{(E-\mu_2)/kT_2}-1}\right] dE
    \label{eq:reciprocal_QW}
\end{align}

\subsection{Nonreciprocal Intermediate Layer}

In the nonreciprocal case, Kirchhoff's law is broken within $[E_g, E_c]$ over a narrow angular range, allowing independent control of emissivity and absorptivity. Parabolic mirrors can be used to focus all emitted and absorbed photons into this narrow angular range from the LED and absorber sides (Fig.~\ref{fig:schematics}). Energy conservation still constrains the optical properties on each side of the filter. On the LED-facing side:
\begin{align}
    \alpha_{\mathrm{LED}} + t_{\mathrm{LED}} + r_{\mathrm{LED}} &= 1 \nonumber\\
    \epsilon_{\mathrm{LED}} + t_{\mathrm{PV}} + r_{\mathrm{LED}} &= 1
\end{align}
where the first equation balances incoming radiation from the LED side and the second balances outgoing radiation from the same side. An analogous pair holds on the PV-facing side:
\begin{align}
    \alpha_{\mathrm{PV}} + t_{\mathrm{PV}} + r_{\mathrm{PV}} &= 1 \nonumber\\
    \epsilon_{\mathrm{PV}} + t_{\mathrm{LED}} + r_{\mathrm{PV}} &= 1
\end{align}

We maximize the nonreciprocal contrast by setting $\epsilon_{\mathrm{LED}} = 1$, $\alpha_{\mathrm{LED}} = 0$ on the LED-facing side and $\alpha_{\mathrm{PV}} = 1$, $\epsilon_{\mathrm{PV}} = 0$ on the PV-facing side. Substituting into the energy conservation constraints and requiring all optical coefficients to be non-negative yields a unique set of solutions where $t_{\mathrm{LED}} = 1$ and $t_{\mathrm{PV}} = 0$. Photons from the LED can transmit freely to the PV, but the backward photon flux from the PV is fully absorbed by the intermediate layer rather than reaching the LED. The layer then re-emits thermal radiation toward the LED at its own temperature $T_m < T_2$, thereby functioning as a radiative heat shield.

Outside $[E_g, E_c]$, the filter is perfectly reflecting on both sides ($r_{\mathrm{LED}} = r_{\mathrm{PV}} = 1$) like the reciprocal filter, which forces $\alpha = \epsilon = t = 0$ by energy conservation. This precludes heat transfer to the intermediate layer at above-cutoff frequencies while also preventing higher-energy photons from transmitting and degrading efficiency and cooling power density.

\subsection{Intermediate Layer Temperature}

Since $t_{\mathrm{PV}} = 0$, the absorber no longer exchanges energy directly with the LED through the filter. Instead, the intermediate layer absorbs radiation from the PV within $[E_g, E_c]$ and re-emits as a thermal body. Modeling the broadband emission and absorption of the layer with the ambient environment at $T_{\mathrm{amb}}$, the energy balance yields
\begin{align}
    \sigma T^4_m &= \sigma T^4_{\mathrm{amb}} + \frac{2\pi}{h^3 c^2}\int_{E_g}^{E_c} \frac{E^3}{e^{E/kT_2}-1}\, dE \nonumber \\
    T_m &= \left[T^4_{\mathrm{amb}} + \frac{2\pi}{\sigma h^3 c^2}\int_{E_g}^{E_c} \frac{E^3}{e^{E/kT_2}-1}\, dE\right]^{1/4}
    \label{eq:Tm_BB}
\end{align}
where $T_{\mathrm{amb}} = 263$~K. Because $E_c = qV_1/(1-T_1/T_2)$ depends on the LED voltage, $T_m$ varies at each bias point. For voltages such that $E_c < E_g$, the PV emits no photons within the permitted frequency range, and we set $T_m = T_{\mathrm{amb}}$. The performance metrics are then given by Eq.~\eqref{eq:reciprocal_QW} with $T_2$ replaced by $T_m$ and $\mu_2 = 0$:
\begin{align}
    Q_L &= \frac{2\pi}{h^3 c^2} \int^{E_c}_{E_g} (E-qV_1) \left[\frac{E^2}{e^{(E-qV_1)/kT_1}-1} - \frac{E^2}{e^{E/kT_m}-1}\right] dE \nonumber\\
    Q_H &= \frac{2\pi}{h^3c^2} \int^{E_c}_{E_g} E\left[\frac{E^2}{e^{(E-qV_1)/kT_1}-1} - \frac{E^2}{e^{E/kT_m}-1}\right] dE \nonumber\\
    W_{\mathrm{in}} &= \frac{2\pi q V_1}{h^3 c^2} \int^{E_c}_{E_g}\left[\frac{E^2}{e^{(E-qV_1)/kT_1}-1} - \frac{E^2}{e^{E/kT_m}-1}\right] dE
    \label{eq:nonrecip_QW_BB}
\end{align}

\subsection{Thermophotonic Configuration with PV Recovery}

\begin{figure}[ht]
    \centering
    \includegraphics[width = 6.5in]{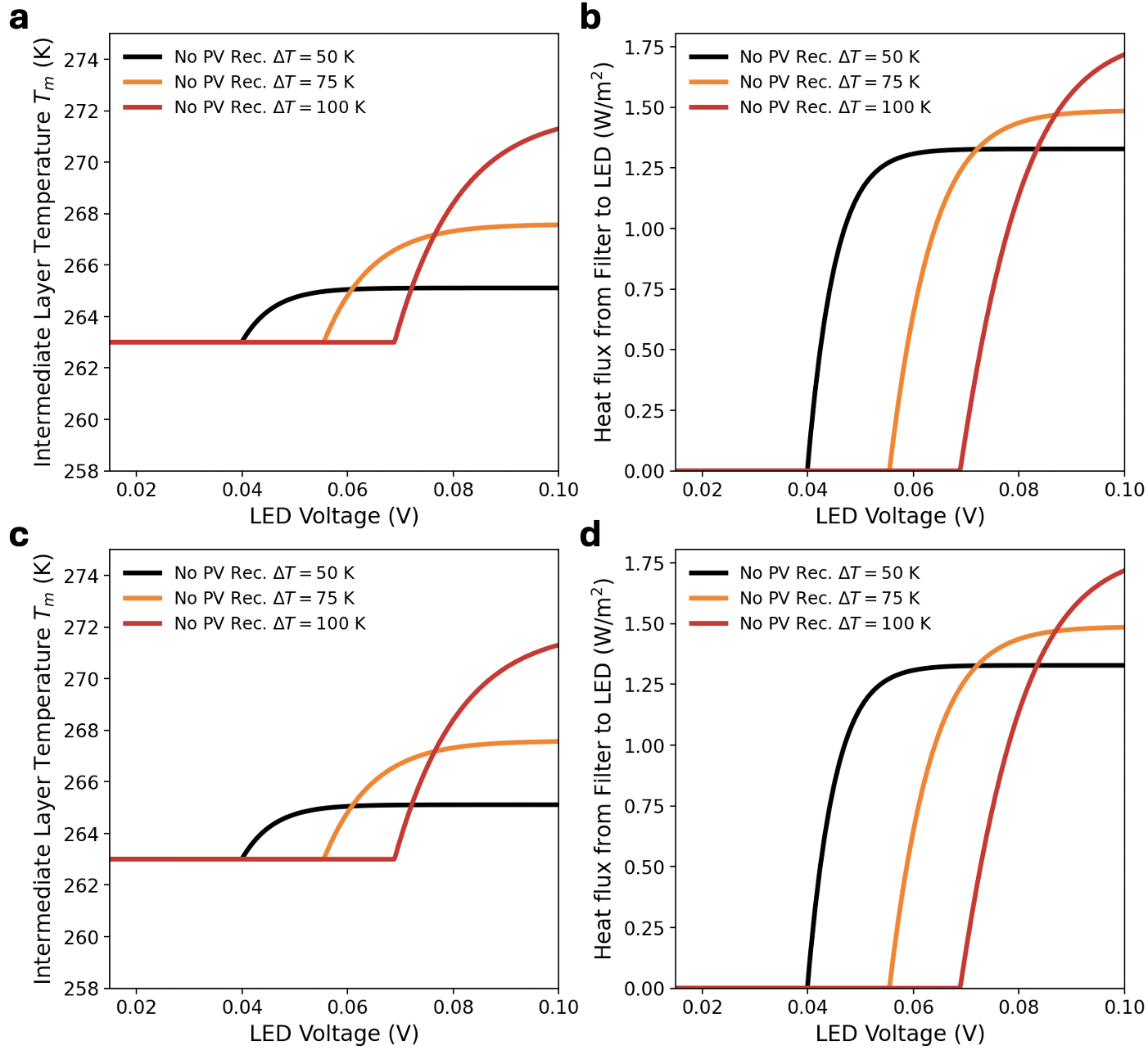}
    \caption{(a),(b) Temperature and heat flux of the nonreciprocal intermediate layer vs LED bias voltage when the hot side is a blackbody absorber, for $\Delta T = 50$~K, 75~K, and 100~K. The cold side temperature is held at 263~K while the hot side temperature is swept. No power recovery occurs in this case. (c),(d) Temperature and heat flux of the intermediate layer when the hot side is a biased PV whose generated power is partially recovered by the LED. Due to the nonzero chemical potential of the biased PV, the intermediate layer absorbs a higher heat flux from the hot side.}
    \label{fig:B_field}
\end{figure}

We now introduce the thermophotonic configuration in which the absorber is replaced by a PV cell and a portion of the generated power is fed back to the LED (Fig.~\ref{fig:schematics}). The PV emits with a nonzero chemical potential $\mu_2 = qV_2$, so the photon flux from the PV side becomes $\phi(E, qV_2, T_2)$. By current conservation in the absence of nonradiative recombination, $J_2 = -J_1$, and the net input power to the system is
\begin{align}
    W_{\mathrm{in}} = J_1 V_1 + J_2 V_2 = J_1(V_1 - V_2)
\end{align}
The cooling power density and COP become
\begin{align}
    Q_L &= \frac{2\pi}{h^3 c^2} \int^{\infty}_{E_g} (E-qV_1)\left[\frac{E^2}{e^{(E-qV_1)/kT_1}-1} - \frac{E^2}{e^{(E-qV_2)/kT_2}-1}\right] dE \\[6pt]
    \mathrm{COP} &= \frac{Q_L}{J_1(V_1 - V_2)}
    \label{eq:COP_TPX}
\end{align}

In the presence of an intermediate filter layer, the integration limits are replaced by $[E_g, E_c]$ with the general cutoff energy from Eq.~\eqref{eq:cutoff_general}. For the nonreciprocal case, the intermediate layer temperature is modified by the nonzero PV chemical potential:
\begin{align}
    T_m = \left[T^4_{\mathrm{amb}} + \frac{2\pi}{\sigma h^3 c^2}\int_{E_g}^{E_c} \frac{E^3}{e^{(E-qV_2)/kT_2}-1}\, dE\right]^{1/4}
    \label{eq:Tm_PV}
\end{align}
and $T_2$ is replaced by $T_m$ in the expressions for $Q_L$, $Q_H$, and $W_{\mathrm{in}}$ as in Eq.~\eqref{eq:nonrecip_QW_BB}. The PV voltage is optimized to maximize the COP in the absence of the filter layer, and the intermediate layer temperature is then solved self-consistently using this optimized voltage.

\begin{figure}[ht]
    \centering
    \includegraphics[width = 6.5in]{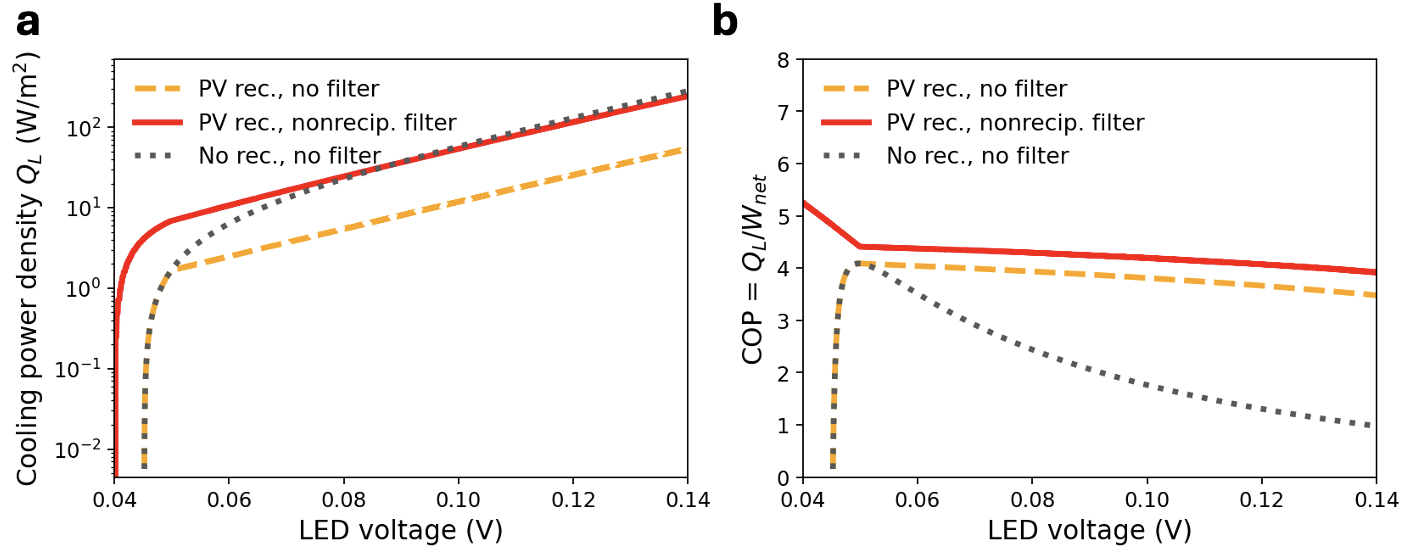}
    \caption{(a),(b) Cooling power density $Q_L$ and COP vs LED bias voltage for three scenarios at $\Delta T = 50$~K. Without a filter, PV recovery improves the COP but reduces $Q_L$ due to the nonzero PV chemical potential. The nonreciprocal filter mitigates backward emission from the hot PV, recovering the cooling power density while maintaining the COP benefit.}
    \label{fig:none_vs_nonrecip}
\end{figure}

In Figs.~\ref{fig:B_field}a and \ref{fig:B_field}b, we show the temperature and heat flux toward the LED side of the nonreciprocal intermediate layer vs LED bias voltage at $\Delta T = 50$~K, 75~K, and 100~K in the absence of PV recovery. The $\Delta T$ values are set by holding the LED side temperature constant at 263~K while increasing the absorber temperature. Because the absorber has zero chemical potential, the temperature and heat flux plateau beyond a certain LED voltage, with the plateaus reaching higher values at larger $\Delta T$. The onset of the temperature and heat flux increase corresponds to the threshold voltage from Eq.~(7), which shifts to higher voltages with greater $\Delta T$. In contrast, Figs.~\ref{fig:B_field}c and \ref{fig:B_field}d show the corresponding quantities with PV recovery at the optimal PV voltage. With a nonzero chemical potential now present on the absorber side, the temperature and heat flux increase much more rapidly than in the passive blackbody case, with the increase being more pronounced at larger $\Delta T$. The sharp turn-on is also absent due to the nonzero PV chemical potential.

Nevertheless, the nonreciprocal intermediate layer simultaneously improves both the cooling power density and COP by residing at an intermediate temperature and emitting less heat toward the LED. In Figs.~\ref{fig:none_vs_nonrecip}a and \ref{fig:none_vs_nonrecip}b, we compare $Q_L$ and COP vs LED voltage across three cases at a fixed $\Delta T = 50$~K: (1) no PV recovery with no filter, (2) PV recovery with no filter, and (3) PV recovery with the nonreciprocal filter. Comparing cases (1) and (2) isolates the tradeoffs of PV recovery: the nonzero PV chemical potential introduces additional parasitic backward emission that decreases the cooling power density (Fig.~\ref{fig:none_vs_nonrecip}a) but reduces the net work input, yielding a higher COP (Fig.~\ref{fig:none_vs_nonrecip}b). With the nonreciprocal filter (case 3), the cooling power density is improved by nearly a factor of 10 by blocking the backward radiative flux from the hot PV (Fig.~\ref{fig:none_vs_nonrecip}a), while maintaining a slight COP benefit (Fig.~\ref{fig:none_vs_nonrecip}b). The reciprocal intermediate layer has a negligible effect on $Q_L$ and COP due to the narrow-band nature of LED emission; noticeable improvements appear only at LED voltages within a few mV of the threshold voltage. These effects are detailed in Figs.~S1 and S2 for the no-recovery and PV-recovery cases, respectively.

\section{Impacts of Nonreciprocity on Realistic TPX Device Models}
 
We next consider a full device model consisting of a GaAs LED on the cold side ($T_C = 290$~K) emitting to a GaAs PV cell on the hot side ($T_H = 310$~K). Following \cite{xiao_electroluminescent_2018}, the relevant performance metrics including parasitic loss channels are
\begin{align}
    W_{\mathrm{in}} &= J_C V_C + Q_{\Omega,C} - (J_H V_H - Q_{\Omega,H}) \nonumber\\
    Q_L &= \bar{E}_C \phi_C - J_C V_C - Q_{\Omega,C} - \bar{E}_H \phi_H - Q_{\mathrm{leak}} \nonumber\\
    Q_H &= \bar{E}_C \phi_C - \bar{E}_H \phi_H - Q_{\mathrm{leak}}
\end{align}
where $Q_{\Omega,C}$ and $Q_{\Omega,H}$ are the Ohmic dissipation at the contacts of the LED and PV respectively, and $Q_{\mathrm{leak}}$ is the parasitic nonluminescent heat leakage from the LED to the PV. For simplicity, we set all parasitic terms to zero ($Q_{\Omega,C} = Q_{\Omega,H} = Q_{\mathrm{leak}} = 0$), so that $W_{\mathrm{in}} = J_C V_C - J_H V_H$, $Q_L = \bar{E}_C \phi_C - J_C V_C - \bar{E}_H \phi_H$, and $Q_H = \bar{E}_C \phi_C - \bar{E}_H \phi_H$, recovering the same structure as the idealized expressions in the previous section.
 
The key difference from the idealized model lies in the treatment of the luminescent photon flux $\phi$ and current density $J$. By detailed balance, the luminescent photon flux from a semiconductor surface is
\begin{align}
    \phi = \int_{E_g}^{\infty} \int_{0}^{\pi/2} \frac{2E^2}{c^2 h^3} \frac{a(E,\theta)}{e^{(E-qV)/kT}-1}\, 2\pi \sin\theta \cos\theta\, d\theta\, dE
\end{align}
We assume that the front surface of the semiconductor is textured with an antireflection coating such that emitted photons trace randomized paths, rendering the absorptivity $a(E,\theta)$ angle-independent. Performing the angular integration then yields
\begin{align}
    \phi = \int_{E_g}^{\infty} \frac{2\pi E^2}{c^2 h^3} \frac{a(E)}{e^{(E-qV)/kT}-1}\, dE
\end{align}
where $\phi_C$ and $\phi_H$ denote the photon fluxes from the cold-side LED and hot-side PV, respectively. The current densities on each side are then
\begin{align}
    J_C &= q\!\left(\frac{\phi_C}{\eta_{\mathrm{ext},C}} - \eta_{\mathrm{abs},C}\,\phi_H\right) \nonumber\\
    J_H &= -q\!\left(\frac{\phi_H}{\eta_{\mathrm{ext},H}} - \eta_{\mathrm{abs},H}\,\phi_C\right)
\end{align}
where $\eta_{\mathrm{ext}}$ and $\eta_{\mathrm{abs}}$ are the external luminescence and absorption quantum efficiencies. We assume $\eta_{\mathrm{abs}} = 1$ on both sides and focus on the impact of nonidealities on $\eta_{\mathrm{ext}}$, which can be decomposed as $\eta_{\mathrm{ext}} = \eta_{\mathrm{int}}\, C_{\mathrm{ext}}$, the product of the internal quantum efficiency and the light extraction efficiency. We further assume $C_{\mathrm{ext}} = 1$ for simplicity.
 
The internal quantum efficiency is
\begin{align}
    \eta_{\mathrm{int}} = \frac{J_{\mathrm{rad}}}{J_{\mathrm{rad}} + J_{\mathrm{SRH}} + J_{\mathrm{Auger}}}
\end{align}
where the radiative recombination current density is given by the van Roosbroeck--Shockley relation:
\begin{align}
    J_{\mathrm{rad}} = qd \int^{\infty}_{E_g} \frac{8\pi n^2_r E^2}{c^2 h^3} \left[\frac{\alpha(E,V)}{e^{(E-qV)/kT}-1} - \frac{\alpha(E)}{e^{E/kT}-1}\right] dE
\end{align}
Here $d$ is the active region thickness, $n_r$ is the refractive index, and $\alpha(E,V)$ is the bulk interband absorption coefficient, which is generally voltage-dependent but becomes independent of voltage for $qV < E_g - 3kT$. The Auger recombination current density is
\begin{align}
    J_{\mathrm{Auger}} = qd\,(C_n n + C_p p)(np - n_i^2)
\end{align}
where $C_n$ and $C_p$ are the Auger coefficients for the two-electron and two-hole processes and $n_i$ is the intrinsic carrier density. The SRH recombination current density is
\begin{align}
    J_{\mathrm{SRH}} = qd\!\left(\frac{1}{\tau_{\mathrm{SRH}}} + \frac{2S}{d}\right) \frac{np - n_i^2}{n + p + 2n_i} + J_{02}\!\left(e^{qV/2kT} - 1\right)
\end{align}
where $\tau_{\mathrm{SRH}}$ is the bulk SRH recombination lifetime, $S$ is the surface recombination velocity, and $J_{02}$ is the depletion-region saturation current density accounting for SRH recombination on the n-side. The electron and hole concentrations under quasi-equilibrium conditions are
\begin{align}
    n &= N_c(T)\, e^{-(E_c - E_{F_n})/kT} \nonumber\\
    p &= N_v(T)\, e^{-(E_{F_p} - E_v)/kT}
\end{align}
where the Boltzmann approximation to the Fermi--Dirac integral of order 1/2 has been used, valid when the quasi-Fermi levels lie more than $3kT$ from the respective band edges. All material parameters used in the following analysis are listed in Table~1.
\begin{table}[h!]
\centering
\caption{Material properties used for GaAs and InP.}
\label{tab:matprops}
\begin{tabular}{llcc}
\toprule
Parameter & Description & GaAs & InP \\
\midrule
$d$           & Layer thickness                          & 200 nm
              & 200 nm \\
$n_r$         & Refractive index                         & 3.6
              & 3.4 \\
$N_D$         & Doping density                           & $2\times10^{17}\ \mathrm{cm^{-3}}$
              & $3\times10^{17}\ \mathrm{cm^{-3}}$ \\
$E_g$         & Band gap                                 & $1.44\ \mathrm{eV}$
              & $1.36\ \mathrm{eV}$ \\
$N_v$         & Effective density of states in valence band
              & $6.64\times10^{18}\ \mathrm{cm^{-3}}$
              & $9.38\times10^{18}\ \mathrm{cm^{-3}}$ \\
$N_c$         & Effective density of states in conduction band
              & $3.57\times10^{17}\ \mathrm{cm^{-3}}$
              & $4.69\times10^{17}\ \mathrm{cm^{-3}}$ \\
$n_i$         & Intrinsic carrier concentration          & $2.50\times10^{4}\ \mathrm{cm^{-3}}$
              & $1.81\times10^{5}\ \mathrm{cm^{-3}}$ \\
$S$           & Surface recombination velocity           & $1.36\ \mathrm{cm/s}$
              & $46.7\ \mathrm{cm/s}$ \\
$\tau_{\mathrm{SRH}}$
              & SRH lifetime                             & $23.2\ \mu \mathrm{ s}$
              & $0.643\ \mu \mathrm{s}$ \\
$C_n, C_p$    & Auger coefficient                        & $2.55\times10^{-30}\ \mathrm{cm^{6}/s}$
              & $9\times10^{-31}\ \mathrm{cm^{6}/s}$ \\
$J_{02}$      & Saturation current density
              & $196\ \mathrm{fA/cm^{2}}$
              & $466\ \mathrm{fA/cm^{2}}$ \\
$\alpha_0(E)$ & Absorption coefficient                   & piece-wise fit
              & piece-wise fit\\
\bottomrule
\end{tabular}
\end{table}

In Figs. 4a and 4b, the nonreciprocal effect on the COP--$Q_L$ loci is shown for GaAs and InP thermophotonic cooling systems at a fixed $\Delta T = 50$~K. The loci exhibit characteristic curved shapes due to parasitic SRH recombination dominating at low LED bias voltages and Auger recombination dominating at high biases. Consistent with the idealized results in Figs. 3a and 3b, the nonreciprocal filter mitigates parasitic backward thermal emission to the LED side by residing at an intermediate temperature, thereby improving both the COP and cooling power density relative to the no-filter case. The reciprocal filter provides only marginal improvement in the presence of nonidealities as well, again owing to the fundamentally narrow-band nature of LED emission (Fig.~S3).

\begin{figure}[h!]
    \centering
    \includegraphics[width = 5.5in]{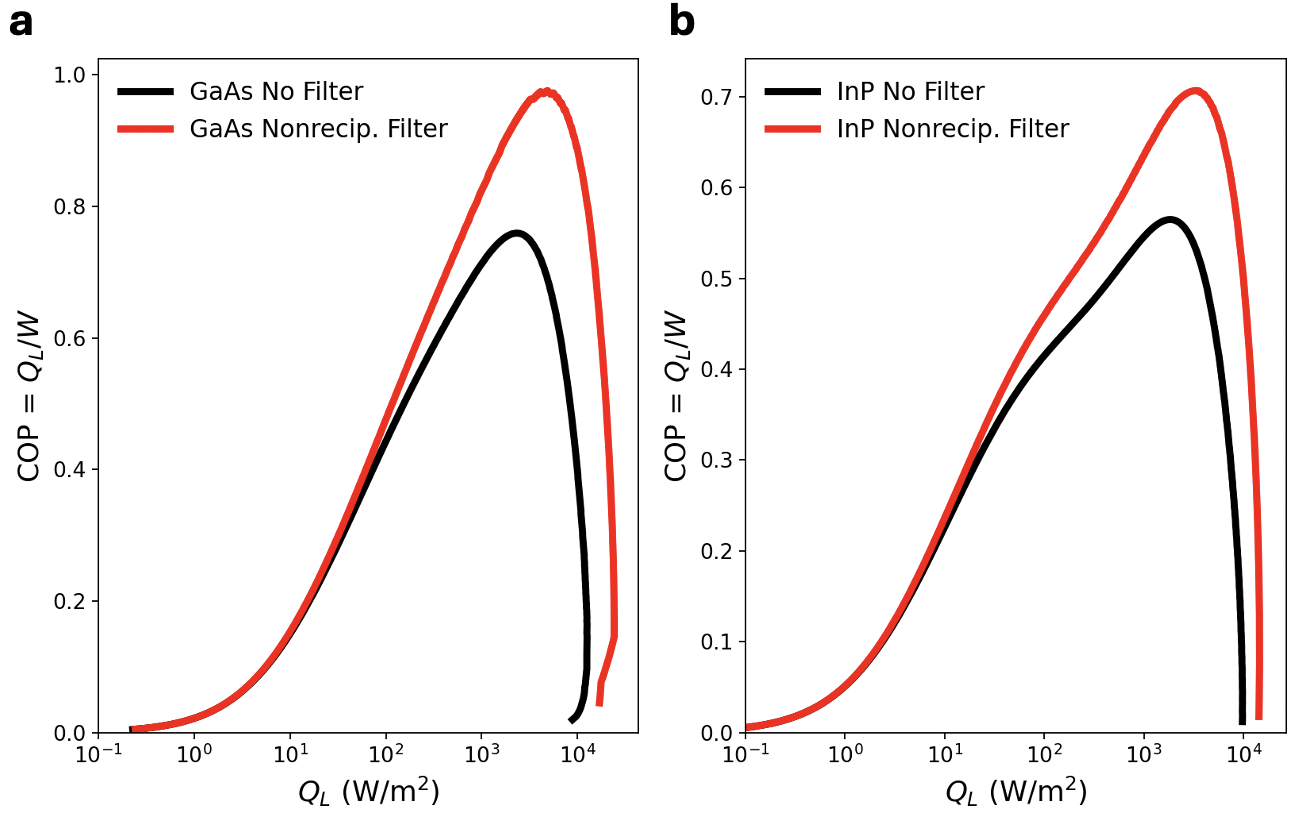}
    \caption{(a),(b) Impact of the nonreciprocal intermediate filter layer on the COP vs cooling power density loci when taking into account nonidealities in GaAs and InP materials. All material parameters and nonradiative recombination coefficients are listed in Table 1. With nonidealities taken into account, the nonreciprocal filter layer  can significantly improve the COP and cooling power density. It should be noted that the $\Delta T$ is fixed at 50 K.} 
    \label{fig:GaAs_InP_dT50K_none_vs_nonrecip}
\end{figure}

To quantify the performance benefit across a range of operating conditions, we show the enhancement in $Q_L$ and COP in Figs. 5a and 5b for GaAs and in Figs. 5d and 5e for InP. In both material systems, the enhancement is greatest in cooling power density across the LED voltage ranges of interest. The COP enhancement, while smaller, is still appreciable and allows the nonreciprocal filter to outperform the no-filter case across a wide range of $\Delta T$ values. The enhancement does decrease slightly with increasing $\Delta T$ as a result of the greater backward radiative flux from the PV and the correspondingly higher temperature of the intermediate nonreciprocal layer. However, as shown in Figs. 5c and 5f for GaAs and InP respectively, operating at higher $\Delta T$ values permits operation closer to the Carnot COP, motivating potential cryogenic cooling applications as noted in prior work.

\begin{figure}[ht]
    \centering
    \includegraphics[width = 6.5in]{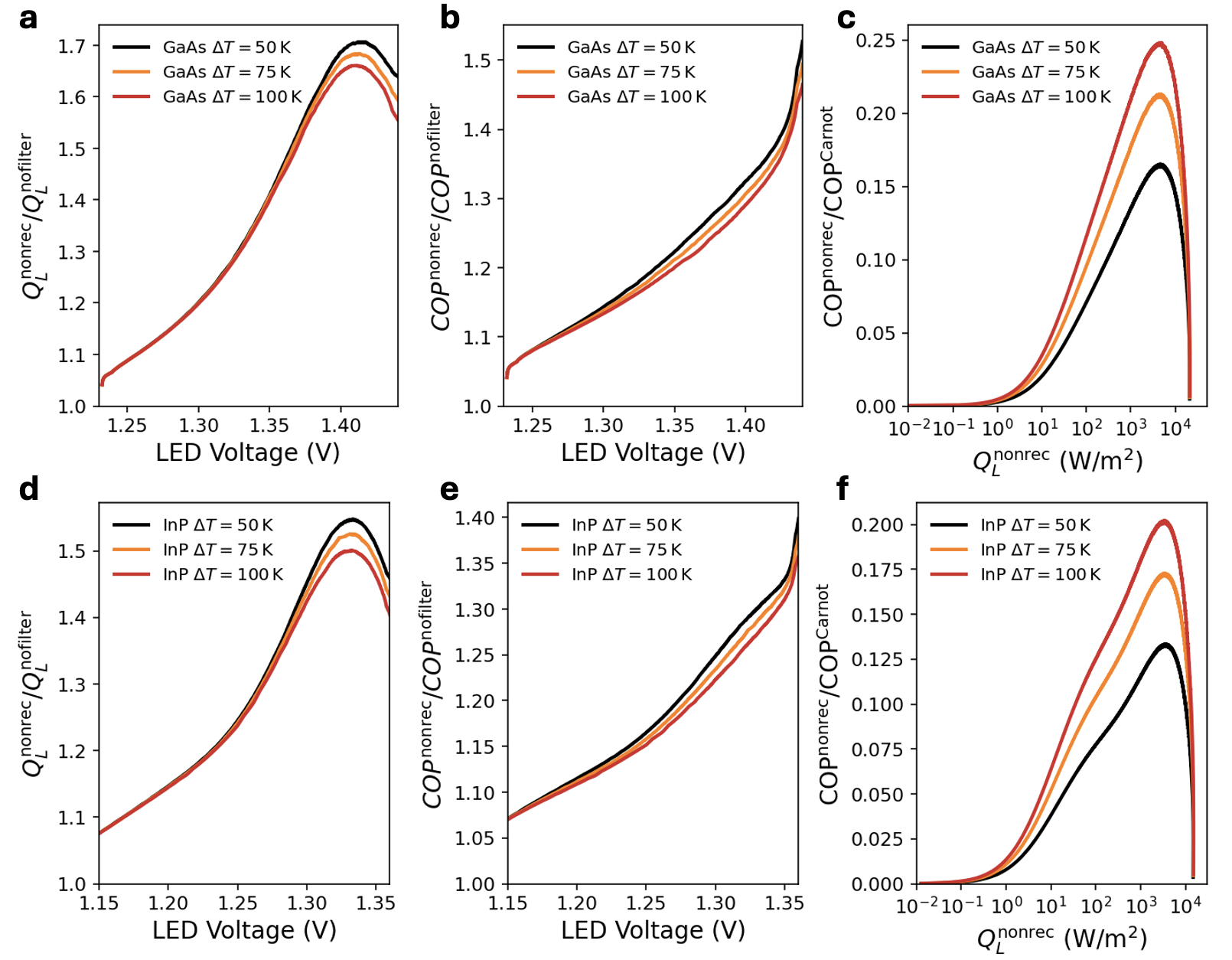}
    \caption{(a),(b),(c) The nonreciprocal enhancement of the cooling power density, enhancement of the COP, and COP as a fraction of the Carnot COP as a function of increasing $\Delta T$'s in a TPX cooling setup composed of GaAs. The cold side temperature is kept fixed at 263K. (d), (e), (f) The same plots with InP as the LED and PV material. With increasing $\Delta T$, the nonreciprocal enhancement of $Q_L$ and COP decreases particularly at higher voltage. However, the system is able to reach closer to the Carnot COP as $\Delta T$ increases. }
    \label{fig:nonrecip_enhancement}
\end{figure}

\section{Conclusion}

In summary, we have shown that while PV recovery in a thermophotonic cooling system can significantly boost the COP, it comes at the cost of reduced cooling power density due to the parasitic backward photon flux from the biased PV cell. By introducing a nonreciprocal intermediate filter layer that violates Kirchhoff's law of thermal radiation, this tradeoff can be circumvented. The nonreciprocal layer absorbs the backward PV flux and re-emits toward the LED at a lower intermediate temperature set by the balance between narrowband PV absorption and broadband ambient exchange, effectively serving as a radiative heat shield. In the idealized case at $\Delta T = 50$~K, the nonreciprocal filter improves the cooling power density by nearly an order of magnitude over the unfiltered TPX case while preserving the COP benefit of PV recovery. We have also shown that these enhancements persist under realistic material nonidealities: incorporating SRH and Auger recombination into GaAs and InP homojunction models yields improvements of approximately 50\% in both cooling power density and COP across temperature differences from 50~K to 100~K. We note that the reciprocal version of the intermediate filter provides only marginal benefit in all cases examined due to the narrow-band nature of LED emission near the band edge.
 
Our analysis assumes maximal nonreciprocal contrast with unity emissivity on one side and unity absorptivity on the other. However, in practice, Kirchhoff-violating materials such as magneto-optical semiconductors under applied fields or Weyl semimetals achieve finite contrast ratios, and examining the sensitivity of the cooling enhancement to partial nonreciprocity would help identify minimum material requirements for practical devices. Our work is also limited to considering far-field radiative exchange; incorporating near-field effects at sub-wavelength gap spacings could further enhance photon flux and thus the achievable cooling power densities. Finally, extending the nonreciprocal filter concept to a multijunction TPX setup \cite{park_multijunction_2024} where lower bandgap junctions emit toward higher bandgap junctions could further improve efficiency, as each junction pair could benefit from its own nonreciprocal intermediate layer to suppress parasitic backward radiation.

\section{Acknowledgments}
This material was based upon work supported by the National Science Foundation under grant no. ECCS-2146577. D.C. acknowledges support of the Department of Defense National Defense Science and Engineering Graduate Fellowship. 

\bibliography{Ref}

\counterwithin{figure}{section}

\clearpage

\end{document}